\documentclass{article}
\usepackage{spconf, amsmath, graphicx}
\usepackage{multirow}

\title{HIGH-QUALITY SPEECH CODING WITH SAMPLE RNN}

\name{Janusz~Klejsa $^{\star}$ \qquad Per~Hedelin $^{\star}$ 
\qquad Cong~Zhou $^{\dagger}$ \qquad Roy~Fejgin $^{\dagger}$
\qquad Lars~Villemoes $^{\star}$}
\address{$^{\star}$ Dolby Sweden AB, Stockholm, Sweden\\
$^{\dagger}$ Dolby Laboratories, San Francisco, CA, USA}

\begin{document}
\ninept
\maketitle
\begin{abstract}
We provide a speech coding scheme employing a generative model based on SampleRNN that, while operating  at significantly lower bitrates, matches or surpasses the perceptual quality of state-of-the-art classic wide-band codecs. Moreover, it is demonstrated that the proposed scheme can provide a meaningful rate-distortion trade-off without retraining. We evaluate the proposed scheme in a series of listening tests and discuss limitations of the approach.
\end{abstract}
\begin{keywords}
speech coding, deep neural networks, vocoder, SampleRNN
\end{keywords}
\section{Introduction}
\label{sec:intro}
Generative modeling for audio based on deep neural networks, such as WaveNet \cite{OordDZSVGKSK16} and SampleRNN \cite{mehri2016samplernn}, has provided significant advances in natural-sounding speech synthesis. The main application has been in the field of text-to-speech \cite{arik2017deep, sotelo2017char2wav, Bonafonte2018} where the models replace the vocoding component. 

Moreover, generative models can be conditioned by global and local latent representations \cite{OordDZSVGKSK16}. In the context of voice conversion, this facilitates natural separation of the conditioning into a static speaker identifier and dynamic linguistic information \cite{Zhou2018}.

In this paper
we develop a wide-band conditioning from the vocoder \cite{Hedelin2000} with parameters quantized to 5.6, 6.4 and 8 kb/s and use these to condition a SampleRNN model. We benchmark our resulting speech coding scheme against AMR-WB \cite{Bessette2002} at 23.05 kb/s and SILK \cite{jensen2010silk} at 16~kb/s, which, at this bitrate, is a state-of-the-art classic speech codec providing good to excellent quality \cite{ramo2010voice}.
 
While it was shown in \cite{Kleijn2018} that the subjective quality of AMR-WB at 23.05~kb/s can be approached at only 2.4~kb/s by using WaveNet synthesis conditioned on narrowband vocoder parameters, the quality gap to classic speech coding schemes was not closed.
Furthermore, the reconstructed waveforms occasionally suffered from typical blind bandwidth extension artifacts (the WaveNet-based decoder used narrow-band conditioning but synthesized wide-band speech).

The purpose of this paper is two-fold. First, we demonstrate that the approach of \cite{Kleijn2018} is reproducible with another generative network architecture. Second, we investigate the scalability of such a scheme in terms of rate-quality trade-off. In particular, we demonstrate that higher reconstruction quality can be achieved by allowing a higher bitrate for the conditioning. The subjective quality is evaluated with a methodology based on MUSHRA\cite{bs.1534-3}. 
 
\par
\begin{figure}[b]
\centering
\hspace{-0.3cm}
\vspace{-0.2cm}
\includegraphics[scale=0.8400]{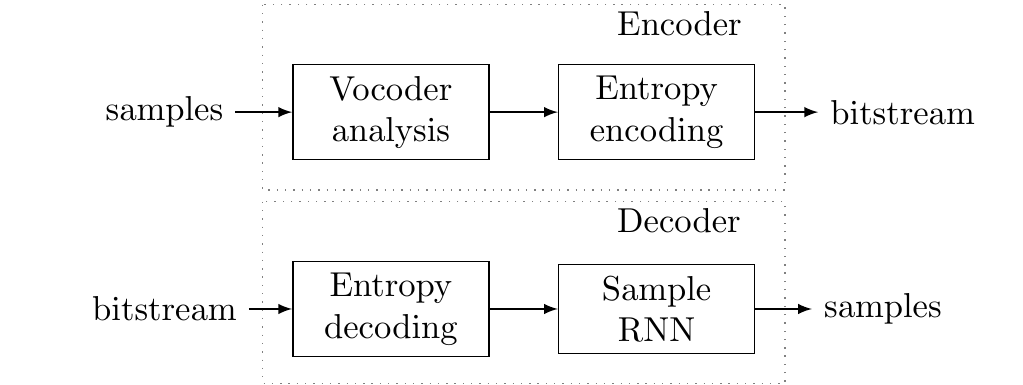}
\caption{\label{fig:block_diagram}Block diagrams of the vocoder-based encoder and the SampleRNN decoder.}
\end{figure}

 The speech coder structure considered in this paper is shown in Fig.~\ref{fig:block_diagram} at a high level. It includes an encoder based on a vocoding structure governed by a parametric signal model that facilitates quantization constrained by bitrate. The decoder scheme uses a four-tier SampleRNN architecture. The conditioning parameters are designed in a way that lower quality conditioning can be embedded into higher quality conditioning, which allows for fixed dimensionality of the conditioning irrespectively of the operating bitrate. The signal sample modeling and synthesis are based on a discretized logistic mixture \cite{Salimans2017} instead of the $8$-bit $\mu$-law domain used in \cite{OordDZSVGKSK16, Kleijn2018}. 

\par Perhaps one of the most interesting questions related to the proposed scheme is whether its performance generalizes to unseen speech signals. While it is relatively straightforward to avoid model overfitting within the selected dataset by using a state-of-the-art training approach (e.g., \cite{ng2016nuts, Goodfellow-et-al-2016}) the issue of robustness of the coding algorithm remains unclear. In order to facilitate comparison to \cite{Kleijn2018} we carried our experiments with the same multi-speaker dataset, namely the Wall Street Journal set WSJ0 \cite{wsj0}. However, it comprises only American English, and seems to have been created with capture devices of limited quality. Hence, we cross-validated the performance of the proposed approach on another publicly available dataset (clean speech from the VCTK corpus \cite{valentini2017noisy}). We demonstrate how the performance of the coding scheme degrades on signals from the new data set and then improves again with retraining on an extended dataset.

This paper is structured as follows. First, the vocoder and its quantization scheme are described in Section \ref{sec:vocoder} along with the embedding approach. Next, we describe our SampleRNN model and provide the details of its vocoder conditioning in Section \ref{sec:samplernn}. The experimental setup and the listening test results are presented in Section \ref{sec:experiments}. Finally, conclusions are presented in Section \ref{sec:conclusions}. 
\vspace{-0.2 cm}
\section{Vocoder with quantized parameters}
\label{sec:vocoder}
The encoder scheme is based on a wide-band version of a linear prediction coding (LPC) vocoder \cite{Hedelin2000}. Signal analysis is performed on a per-frame basis, and it results in the following parameters: \emph{i)} an $M$-th order LPC filter, \emph{ii)} an LPC residual RMS level $s$, \emph{iii)} pitch $f_0$, and, \emph{iv)} a $k$-band voicing vector $v$. A voicing component $v(i)$, $i=1,\dots,k$ gives the fraction of periodic energy within a band. All these parameters are used for conditioning of SampleRNN, as described in Section \ref{sec:samplernn}. We note the signal model used by the encoder aims at describing only clean speech  (without background or simultaneously active talkers).
\par The analysis scheme operates on 10~ms frames of a signal sampled at 16~kHz. In the proposed encoder design, the order of the LPC model, $M$, depends on the operating bitrate. Standard combinations of source coding techniques are utilized to achieve encoding efficiency with appropriate perceptual consideration, including vector quantization (VQ), predictive coding and entropy coding \cite{gersho2012vector}. In this paper, for all experiments, we define the operating points of the encoder as in Table \ref{tab:voc1}. We used standard tuning practices. For example, the spectral distortion for the reconstructed LPC coefficients is kept close to $1$~dB \cite{paliwal1990efficient}.
\par The LPC model is coded in the line spectral pairs (LSP) domain utilizing prediction and entropy coding. For each LPC order, $M$, a Gaussian mixture model (GMM) was trained on the WSJ0 train set, providing probabilities for the quantization cells. Each GMM component has a $\mathcal{Z}$-lattice according to the principle of union of $\mathcal{Z}$-lattices \cite{shabestary2003entropy, shabestary2005vector}. The final choice of quantization cell is according to a rate-distortion weighted criterion.

\begin{table}[t]
\centering
\caption{\label{tab:voc1} Operating points of the encoder ($k=6$)}
\begin{tabular}{c|ccccc}
\hline
 $r_{\text{nominal}}$ &  $M$ & spectral & $n$ bits & $n$ bits \\ 
 $\text{[kb/s]}$   & & dist. $\text{[dB]}$ & $s$ & $v_w$ \\ \hline
 8.0  & 22 & 0.754 & 1 + 9 & 9 \\
 6.4   & 16 & 0.782 & 1 + 8 & 9 \\
 5.6 & 16 & 1.33 & 1 + 8 & 9 \\ \hline
\end{tabular}
\end{table}

\par The residual level $s$ is quantized in the dB domain using a hybrid approach similar to that in \cite{linden1996improving}. Small level inter-frame variations are detected, signalled by one bit, and coded by a predictive scheme using fine uniform quantization. In other cases the coding is memoryless with a larger, yet uniform, step-size covering a wide range of levels. 
\par Similar to level, pitch is quantized using a hybrid approach of predictive and memoryless coding. Uniform quantization is employed but executed in a warped pitch domain. Pitch is warped by $f_w = c f_0 /(c+f_0)$ where $c = 500$~Hz and $f_w$ is quantized and coded using $10$ bit/frame.
\par Voicing is coded by memoryless VQ in a warped domain. Each voicing component is warped by $v_w(i) = \log(\frac{1-v(i)}{1+v(i)})$.  A $9$ bit VQ was trained in the warped domain on the WSJ0 train set. 
 
\par A feature vector $\mathbf{h_f}$ for conditioning SampleRNN is constructed as follows. The quantized LPC coefficients are converted to reflection coefficients. The vector of refection coefficients is concatenated with the other quantized parameters, i.e. $f_0$, $s$, and $v$. In the remainder of the paper we use either of two constructions of the conditioning vector. The first construction is the straightforward concatenation described above. For example, for $M=16$, the total dimension of the vector $\mathbf{h_f}$ is $24$; for $M=22$ it is $30$. The second construction is an embedding of lower-rate conditioning into a higher-rate format. For example, for $M=16$, an $22$-dimensional vector of the reflection coefficients is constructed by padding the $16$ coefficients with $6$ zeros. The remaining parameters are replaced with their coarsely quantized (low bitrate) versions, which is possible since their locations within $\mathbf{h_f}$ are now fixed.
\vspace{-0.2 cm}
\section{Conditional Sample RNN}
\label{sec:samplernn}
\begin{figure}
\centering
\includegraphics[scale=0.65]{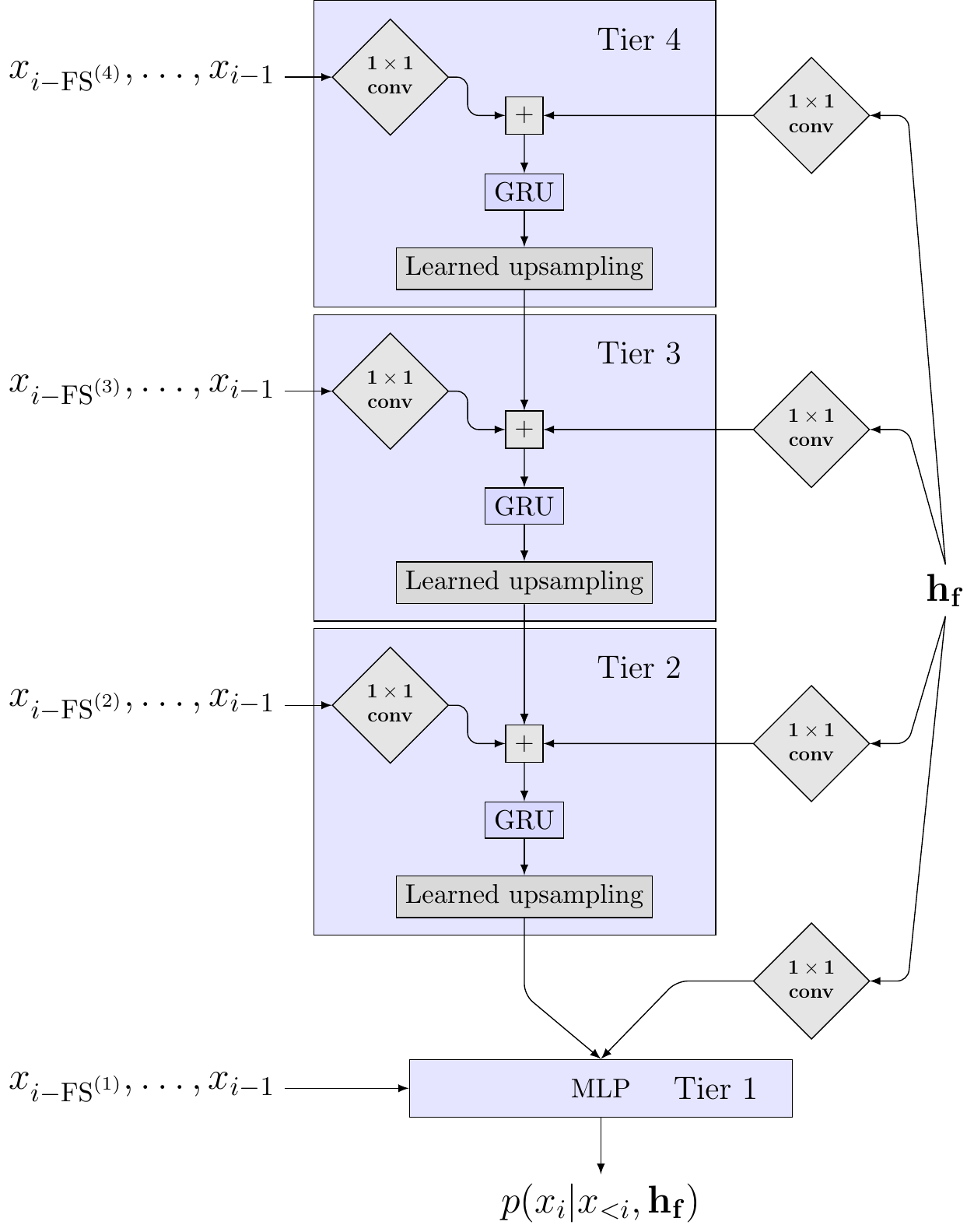}
\caption{\label{fig:cond_sRNN}Conditioning of the SampleRNN.}
\end{figure}
SampleRNN is a deep neural generative model proposed in \cite{mehri2016samplernn} for generating raw audio signals. It consists of a series of multi-rate recurrent layers, which are capable of modeling the dynamics of a sequence at different time scales. SampleRNN models the probability of a sequence of audio samples via factorization of the joint distribution into the product of the individual audio sample distributions conditioned on all previous samples. The joint probability distribution of a sequence of waveform samples $X=\{x_1,\dots,x_T\}$ can be written as \begin{equation}
p(X)=\prod_{i=1}^{T}p(x_i|x_1,\dots,x_{i-1}).
\label{eq:px}
\end{equation}
At inference time, the model predicts one sample at a time by randomly sampling from $p(x_i|x_1,\dots,x_{i-1})$. Recursive conditioning is then be performed using the previously reconstructed samples.
\begin{figure*}[t!]
    \begin{minipage}{.485\textwidth} 
        \text{(a)}\hspace{-0.075cm}
        \includegraphics[scale=0.500]{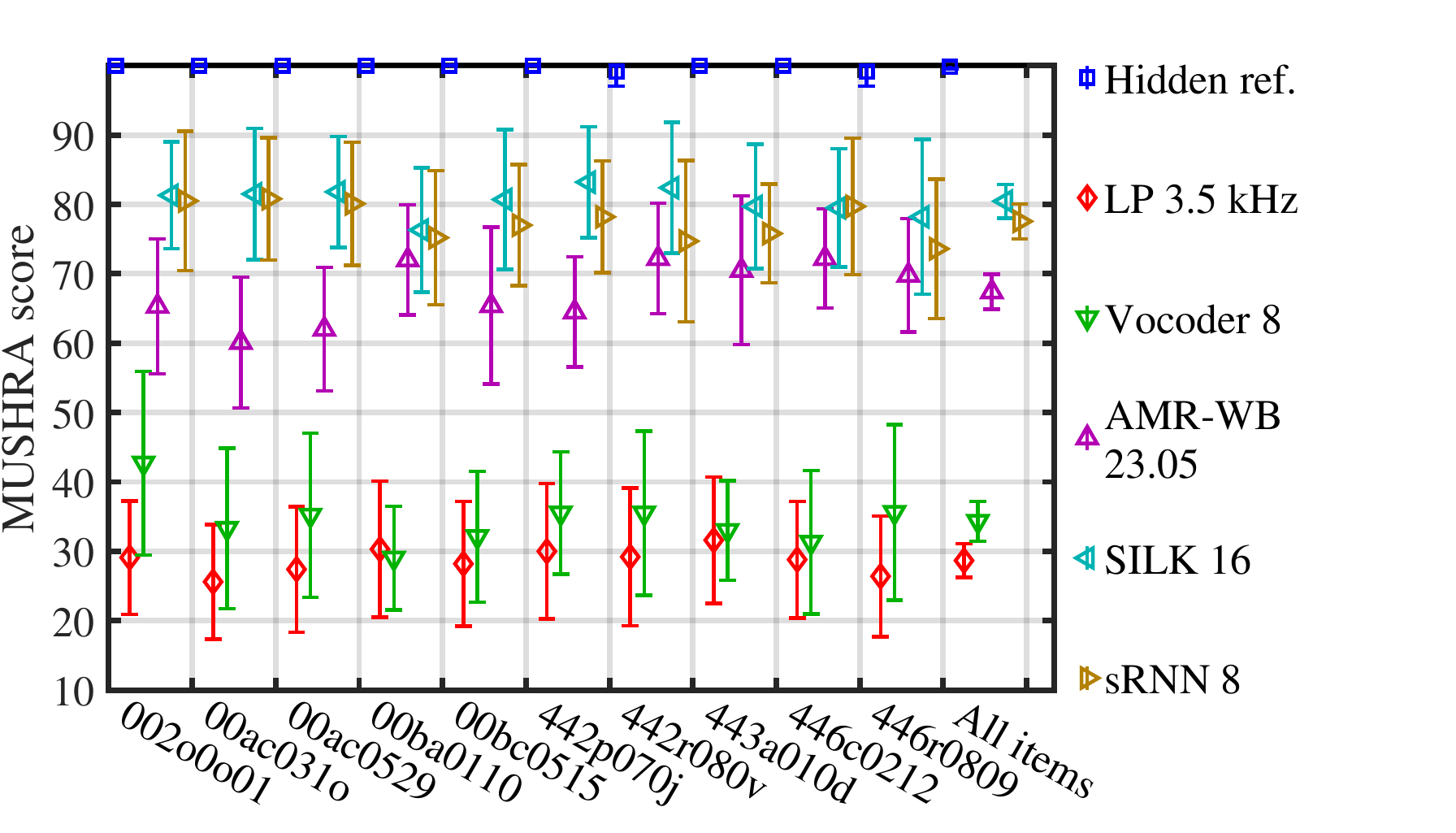}
    \end{minipage}
    \begin{minipage}{.485\textwidth}
        \text{(b)}  \hspace{-0.075cm}
        \includegraphics[scale=0.500]{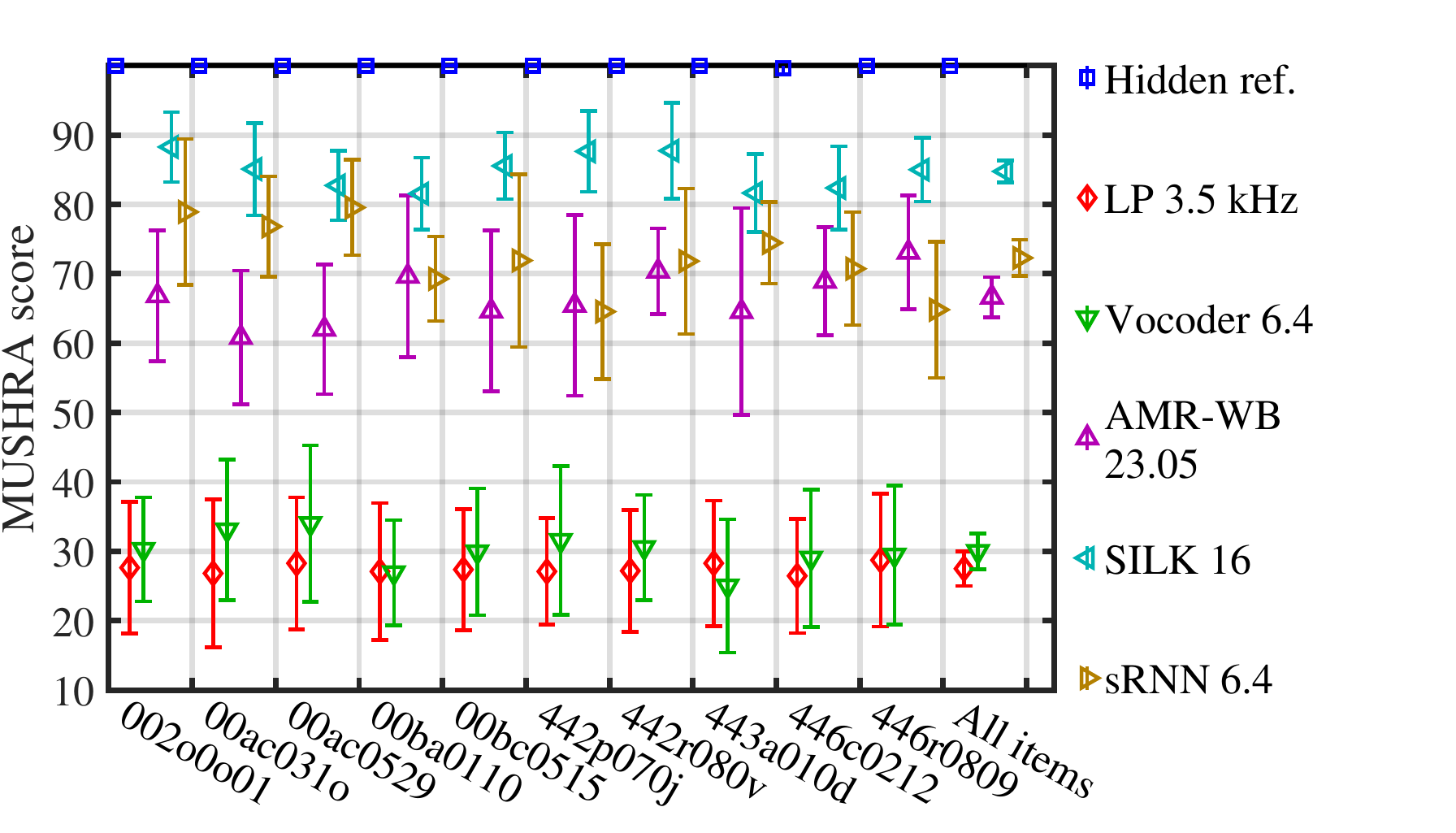}
    \end{minipage}
    \caption{\label{fig:wsj_mushra}Results of the listening test comparing sRNN and vocoder (operating at 8~kb/s (a) and 6.4~kb/s (b)) and the reference speech codecs: AMR-WB (23.05~kb/s) and SILK (16~kb/s); there were 10 listeners in the test (a), and 11 in the test (b); we use original names of the excerpts.}
\end{figure*}
\subsection{Conditioning}
Without conditioning information, SampleRNN is only capable of ``babbling''. Hence, we provide decoded vocoder  parameters, $\mathbf{h_f}$, as conditioning information to the model. Eq.~\ref{eq:px} thus becomes  \begin{equation}
p(X|H)=\prod_{i=1}^{T}p(x_i|x_1,\dots,x_{i-1}, \mathbf{h_f}),
\label{eq:pxh}
\end{equation} where $\mathbf{h_f}$ represents the vocoder parameters corresponding to the audio sample at time $i$.
\par The structure of a conditional SampleRNN is illustrated in Fig.~\ref{fig:cond_sRNN}. In a $K$-tier conditional SampleRNN, the $k$-th tier ($1<k\leq K$) operates on non-overlapping frames of length $\text{FS}^{(k)}$ samples at a time, and the lowest tier ($k=1$) predicts one sample at a time. Waveform samples $x_{i-\text{FS}^{(k)}},\dots,x_{i-1}$ and decoded vocoder conditioning vector $\mathbf{h_f}$ processed by respective $1\times1$ convolution layers are the inputs to $k$-th tier. When $k<K$, the output from ${(k+1)}$-th tier is additional input. Similar to \cite{Zhou2018}, all inputs to $k$-th tier are linearly summed up. The $k$-th RNN tier ($1<k\leq K$) consists of one gated recurrent unit (GRU) layer, cf. \cite{cho2014learning}, and one learned upsampling layer performing temporal resolution alignment between tiers. The lowest ($k=1$) tier consists of a multilayer perceptron (MLP) with $2$ hidden fully connected layers.

\vspace{-0.2cm}
\subsection{Discretized logistic mixture}
Instead of using 256-way softmax in the $8$-bit $\mu$-law domain as in \cite{Bonafonte2018, Kleijn2018}, we adopt the discretized mixture of logistic distributions technique introduced in \cite{Salimans2017, oord2017parallel} to generate 16-bit samples. The output layer of the sample-level MLP predicts parameters of each mixture component, including mixture weight, mean, and log scale. Unlike the original SampleRNN, which performs an embedding of quantized sample values into real-valued vectors, we use the raw samples directly as the input to the MLP tier.

\vspace{-0.2cm}
\subsection{Model configurations and training setup}
As shown in Fig. \ref{fig:cond_sRNN}, our architecture has a four-tier configuration  ($K=4$), where the frame size for the $k$-th tier is $\text{FS}^{(k)}$. We use the following frame sizes: $\text{FS}^{(1)}=\text{FS}^{(2)}=2$, $\text{FS}^{(3)}= 16$ and $\text{FS}^{(4)}=160$. The top tier shares the same temporal resolution as the vocoder parameter conditioning sequence. The learned upsampling layer is implemented through a transposed convolution layer, and the upsampling ratio is $2$, $8$, and $10$ respectively in the second, third and fourth tier. The recurrent layers and fully connected layers contain $1024$ hidden units each.

\par During training, we use a batch size of 24 and a sequence length of 6400 samples for truncated back propagation through time (TBPTT, \cite{williams1990efficient, sutskever2013training}) on a single GPU. We use the ADAM optimizer \cite{kingma2014adam} ($\beta_1=0.9$, $\beta_2=0.999$, and $\epsilon=1\text{e-}{8}$) with an initial learning rate of 0.0002. The learning rate is reduced by multiplying by a factor $0.3$ when validation loss has stopped dropping. Gradients are hard-clipped to a range of $[-1,1]$.

\section{Experimental evaluation}
\label{sec:experiments}
\subsection{Perceptual evaluation on the WSJ0 set}
\label{ssec:mushra_wsj0}
\par We performed perceptual evaluation of the our SampleRNN-based decoder on randomly selected excerpts from the test set of WSJ0 in a listening test resembling the experiment in \cite{Kleijn2018}. There were 5 male speakers and 5 female speakers. The items in the original WSJ0 data set have inconsistent loudness, which is problematic for a listening test. Thus, we performed loudness alignment of the test items prior to encoding. 
\par We used a headphone based MUSHRA-like test methodology \cite{bs.1534-3}. Since the speech data set was only available at 16~kHz sampling frequency, we only included the 3.5~kHz low-pass anchor (LP~3.5~kHz), a wide-band original and hidden reference, and no 7~kHz anchor. The conditions used in the test included SILK (at 16~kb/s, variable rate), which is a state-of-the-art speech codec, and AMR-WB (at 23.05 kb/s, constant bitrate) in a configuration that facilitates comparison of our results to the 2.4~kb/s decoder of \cite{Kleijn2018}. In the first two listening tests, we included pairs of conditions: reconstruction provided by the SampleRNN (sRNN) decoder, and reconstuction provided by the vocoder \cite{Hedelin2000}, which was used to produce the conditioning. Thus both conditions represent decoding of the same bitstream. In the two following experiments, we used sRNN decoders that were trained specifically for their respective operating bitrates. 

\par In the first experiment, we performed a test where the vocoder and the sRNN decoder operated at the average bitrate of 8~kb/s. There were 10 expert listeners in the test. The test results are shown in Fig.~{\ref{fig:wsj_mushra}a, plotted along with the 95\% confidence intervals (normal distribution). It can be seen that the sRNN decoder provides performance comparable to SILK at half of its bitrate. 

\par In a second experiment, we used an average bitrate of 6.4~kb/s for the vocoder and sRNN. The results of the test are shown in Fig.~{\ref{fig:wsj_mushra}b. There were 11 expert listeners taking part in this test. It can be seen that the 6.4 kb/s sRNN  provides performance better than AMR-WB at 23.05~kb/s, but it remains in a similar quality region. However, it can also be seen that sRNN at 6.4 kb/s is significantly worse than SILK. We note that the two tests were carried out in separate sessions. All listeners were experts, and the listener populations were only partially overlapped between the two tests.

\par In addition, we evaluated the performance of the conditions from the listening test using the objective quality assessment tool POLQA \cite{ITUTP863}. The results are shown in Table \ref{tab:polqa}. Although this tool is not suitable for evaluation of signals synthesized by non-deterministic generative models, it is interesting to note a relatively high score for the vocoded conditions, and relatively low scores for sRNN. These results deviate significantly from the results of our listening tests. However, the ordering of the classic codecs, AMR-WB and SILK is the same as in our listening tests.

\begin{table}[b!] 
\vspace{-0.3 cm}
\caption{\label{tab:polqa} Average POLQA scores for conditions and items from the listening tests (wide-band setting)}
\begin{tabular}{ccccccc}
\hline
\textit{} & \multicolumn{2}{c}{Vocoder} & AMR-WB & SILK & \multicolumn{2}{c}{sRNN} \\ \hline
Rate [kb/s]  & 6.4  & 8.0  & 23.05  & 16.0 & 6.4  & 8.0   \\
MOS-LQO   & 3.43  & 3.67  & 4.39  & 4.41 & 3.27  & 3.48 \\  \hline         
\end{tabular}
\end{table}

\vspace{-0.5 cm}
\subsection{Quality-bitrate trade-off}
In a third experiment, we evaluated the perceptual quality-bitrate trade-off achieved with the sRNN decoder. The conditions in this test included sRNN decoders evaluated in Section \ref{ssec:mushra_wsj0}. In addition, we took an 8~kb/s decoder and used it to decode bitstreams generated at lower rates (5.6 kb/s and 6.4~kb/s) -- without retraining. We used properties of the conditioning that allowed to embed a lower rate conditioning into a higher rate conditioning and provide this to a decoder trained on higher-rate conditioning (see Section \ref{sec:vocoder}).
\begin{figure}[t!]
\centering
\includegraphics[scale=0.500]{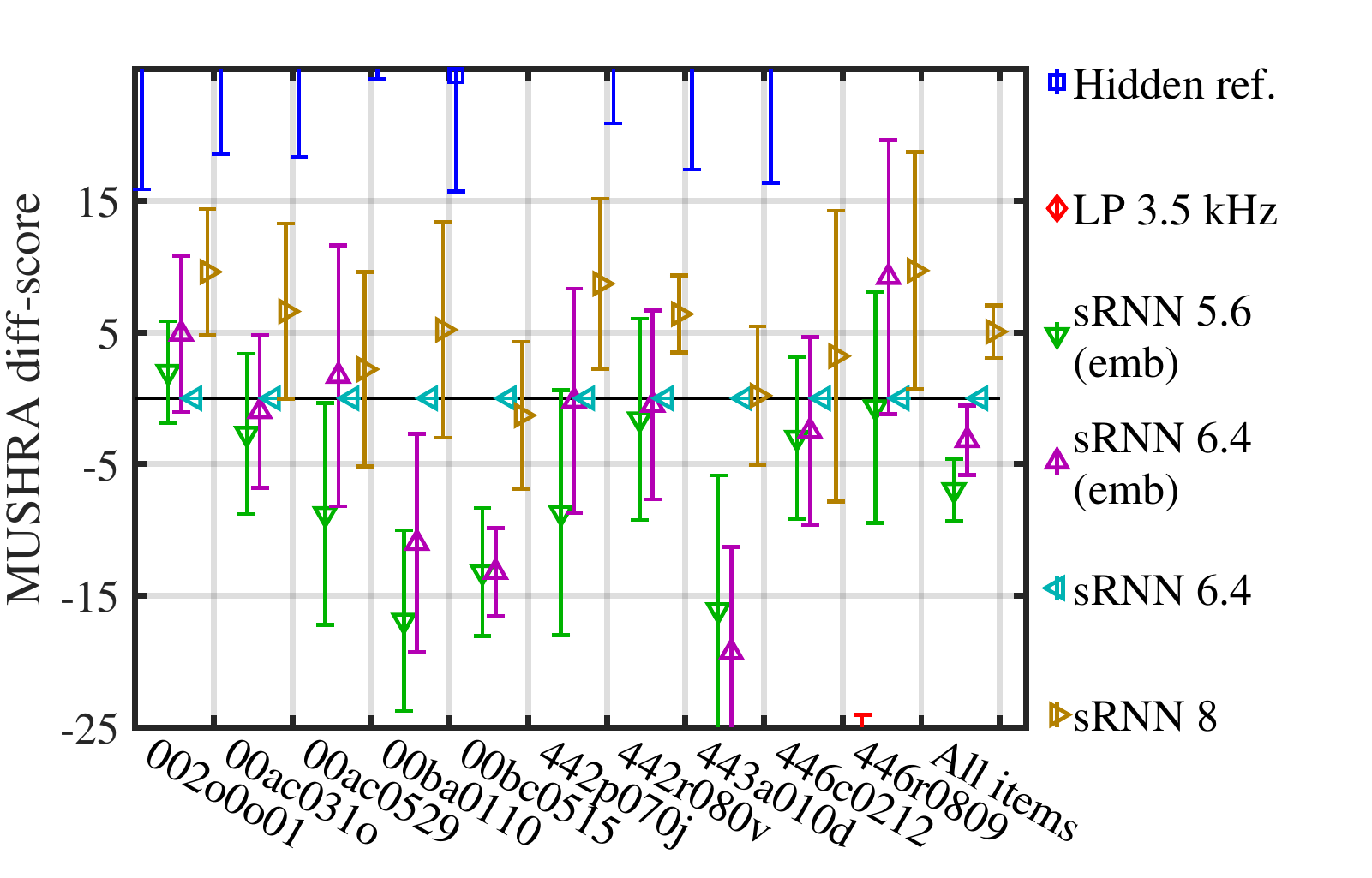}
        \caption{\label{fig:tradeoff}The differential scores in the rate-quality test of the embedded sRNN decoders with respect to the sRNN trained on the 6.4~kb/s conditioning (95\% confidence intervals, Student-t distribution.)}
        \vspace{-0.3 cm}
\end{figure}
\par The results of the experiment are shown in Fig.~\ref{fig:tradeoff}. There were 10 expert listeners in the test. It can be seen that, for the provided data points, sRNN~8 decoder provides a graceful quality degradation with decreasing bit rate (sRNN~6.4 (emb), sRNN~5.6 (emb)). Furthermore, the performance of the decoder using the embedded conditioning (sRNN~6.4 (emb)) is similar to the performance of a decoder trained for the specific bitrate (sRNN~6.4). While we cannot conclude that the result generalizes to a wide range of bitrates, we can see that the proposed sRNN decoder learned to generalize the conditioning, which can be used to provide meaningful rate-distortion trade-offs. We note that in the course of the training the 8~kb/s decoder was never exposed to lower quality conditioning.

\subsection{Perceptual evaluation outside the WSJ0 set}
We found no evidence of model overfitting in our experiments within the WSJ0 dataset and all evaluation was done on recordings not included in training. However, while evaluating the coding scheme on speech signals outside WSJ0 we found a performance drop in order of $10$-$15$ MUSHRA points on average. In order to evaluate this effect in a controlled experiment, we used signals from another publicly available speech database. 
\par We constructed a new dataset based on roughly equal contributions from the WSJ0 set and VCTK set (clean speech from \cite{valentini2017noisy}), and also TSP set \cite{Kabal2002}($1\%$ of the full data set). The final set composition resulted in doubling the total number of speakers. The total recording time of the combined set was still similar to WSJ0 in the previous experiments. The training set comprised $45$ hours of speech and we used $10$ hours of speech for the validation set and $10$ hours for the test set. There was no speaker overlap between the three sets and the material was approximately equally  balanced between female and male speakers. We retrained the sRNN decoder on this new dataset.
\par For our final experiment, we randomly selected items from WSJ0 and VCTK test sets ($4$ female speakers, $4$ male speakers), and used two versions of identically configured sRNN decoders ($8$~kb/s). One was trained solely on WSJ0, the other one was trained on the combination of the datasets as described above. We also used the two classic speech codecs as the quality anchors configured identically as in the previous experiment. The listening test was conducted with 9 expert listeners and its mean results are shown in Fig.~\ref{fig:new_experiment} together with 95\% confidence intervals (normal distribution).
\par It can be seen in the results that the sRNN trained on WSJ0 has significantly worse performance on the items coming from the VCTK set. But the performance of the retrained sRNN is similar to the performance reported in Section~\ref{ssec:mushra_wsj0}. We note that the performance on the WSJ0 items has been maintained while the performance on the signals from the updated data set has improved significantly. The main difference between the WSJ0 items and the VCTK items is that the latter comprise dry speech captured in a semi-anechoic chamber.
\par We have also informally evaluated the scheme on material in Mandarin, French, Swedish, and German and we did not find a performance drop due to operation on speech in these languages. Note that all the training material was in English.

\begin{figure}
\centering
\includegraphics[scale=0.50]{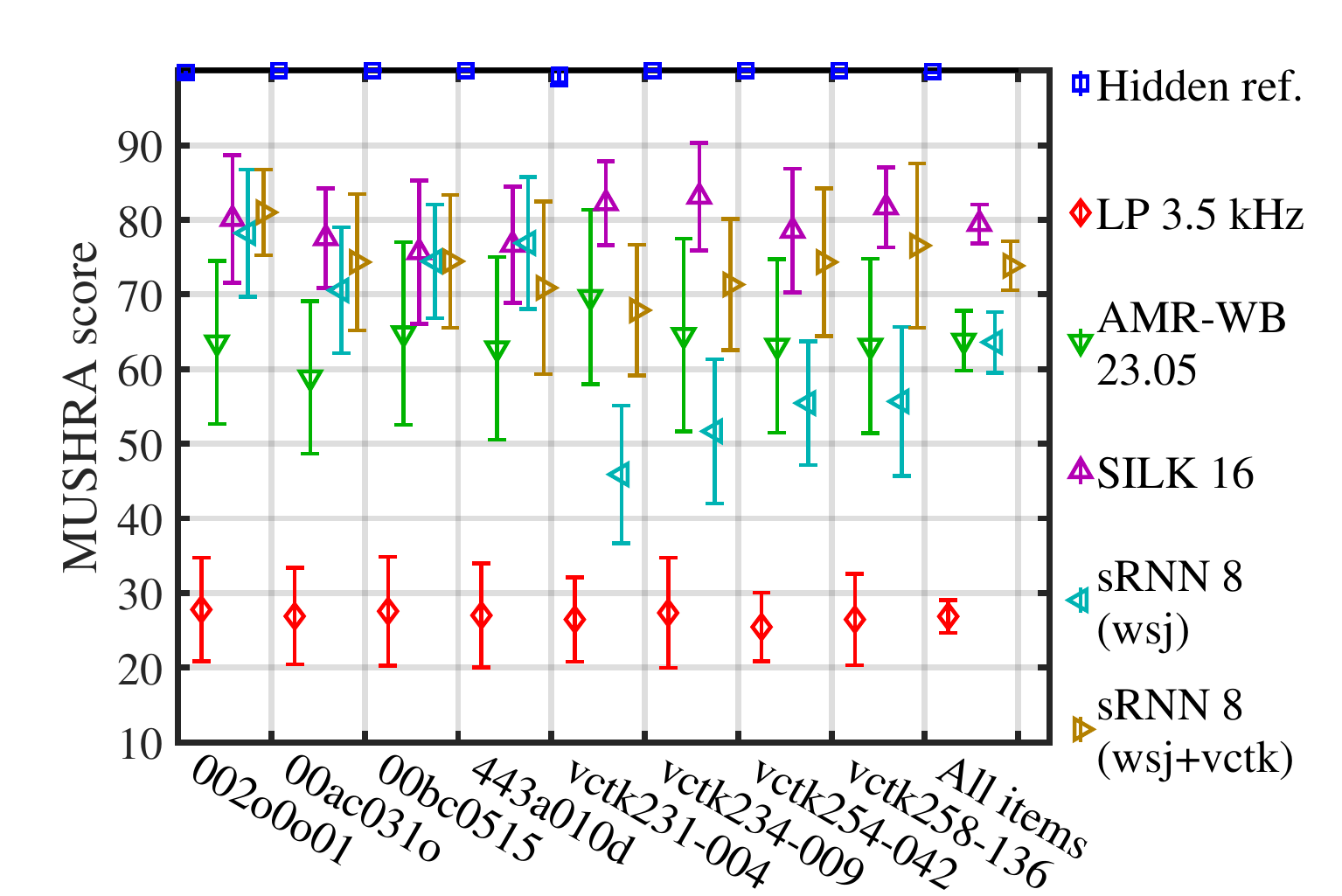}
\caption{\label{fig:new_experiment}The results of listening test including sRNN decoders operating at 8~kb/s trained exclusively on WSJ0 and combination of WSJ0 and VCTK; there were 9~expert listeners.}
\vspace{-0.3 cm}
\end{figure}

\section{Conclusions}
\label{sec:conclusions}
We demonstrated that it is possible to achieve high quality speech coding that is on par with the state-of-the-art speech codecs, but at much lower bitrates. It is possible in a coding scheme, where quantized vocoder parameters are used to condition the SampleRNN. Furthermore, we demonstrated that this architecture has a potential to provide a rate-quality trade-off, which we established by conducting a series of formal listening tests.
\par We found the performance of such schemes is highly dependent on the composition of the datasets and that the robustness issues can be addressed by using more diverse training material.

\vfill\pagebreak


\bibliographystyle{IEEEbib}
\bibliography{refs}

\end{document}